\documentclass{optica-article}
\usepackage[colorinlistoftodos]{todonotes}

\journal{opticajournal} 

\articletype{Research Article}

\usepackage{lineno}

\begin{document}

\title{Impact of nonlinear spectral broadening on the phase noise properties of electro-optic frequency comb}

\author{Aleksandr Razumov,\authormark{1,*} Yijia Cai,\authormark{2}, Jasper Riebesehl,\authormark{1}, Eric Sillekens,\authormark{2}, Ronit Sohanpal,\authormark{2}   Francesco Da Ros,\authormark{1}, Zhixin Liu ,\authormark{2}   and Darko Zibar\authormark{1}}
\address{\authormark{1}Department of Electrical and Photonics Engineering, Technical University of Denmark (DTU), DK-2800 Kgs. Lyngby, Denmark\\
\authormark{2}Optical Networks Group, Department of Electronic and Electrical Engineering, University College London, London, United Kingdom\\}

\email{\authormark{*}alrazu@dtu.dk} 


\begin{abstract*} 
Electro-optic modulation is an attractive approach for generating flat, stable, and low-noise optical frequency combs with relatively high power per comb line. However, a key limitation of electro-optic combs is the restricted number of comb lines imposed by the available RF source power. To overcome this limitation, a nonlinear spectral broadening stage is typically employed.
The phase noise characteristics of an electro-optic comb are well described by the standard phase noise model, which depends on two parameters: the seed laser and the RF source phase noise. A fundamental question that arises is how nonlinear broadening processes affect the phase noise properties of the expanded comb.
To address this, we employ coherent detection, digital signal processing, and subspace tracking.  Our experimental results show that the nonlinearly broadened comb preserves the standard phase noise model of the input electro-optic comb. In other words, the nonlinear processes neither introduce additional phase noise terms nor amplify the existing contributions from the seed laser and RF source. Hence, nonlinear broadening can be viewed as equivalent to driving the electro-optic comb with a much higher RF modulation power. 


\end{abstract*}


\section{Introduction}
Cavity-less optical frequency combs based on electro-optic (EO) modulation have gained significant attention due to their tunable spacing, flat spectral profile, and robustness under diverse operating conditions \cite{Simon}. These combs are particularly attractive for applications such as high-precision spectroscopy, metrology, and optical communications \cite{comb_application,Diddams:10,Leif, Chatterjee:25,Arnaud}. However, compared to cavity-based comb generators (e.g.~modelocked laser, microresonators), the cavity-less approach generally produces fewer comb lines because it lacks cavity enhancement of the nonlinear interaction. A common strategy to extend the number of comb lines is to incorporate a nonlinear mixing stage \cite{Simon,Yijia_apl,Tong:12}. In practice, this typically involves a pulse shaper (or compressor) followed by a nonlinear medium, where parametric processes enable significant spectral broadening \cite{Simon, Yijia_apl}. These types of optical frequency combs are commonly referred to as parametric optical frequency combs.   

Although the phase noise properties of electro-optic combs are well studied and documented \cite{ZhixinEO,Razumov:23, Ishizawa}, detailed investigations of how nonlinear processes affect the phase noise of spectrally broadened combs remain limited. In particular, understanding the contribution of intrinsic nonlinear processes to the total phase noise of the broadened comb is crucial to engineer low-noise combs suitable for spectroscopy, metrology, and coherent optical systems \cite{Lidars, phase_coh_comm}.

One of the challenges in quantifying the impact of the nonlinear process is due to the complicated signal-noise interaction in the nonlinear broadening processes. This includes the interaction between amplitude and phase noise of the seed comb with the nonlinear processes such as self-phase modulation, (SPM), and four wave mixing (FWM). Additionally, the nonlinear interaction depends on the length of nonlinear fiber, dispersion, polarization extinction ratio, pulse-shapes, and repetition rate, making it difficult to derive general conclusions.

Several studies have characterized phase noise in parametric generation using both theoretical modeling and experimental measurements \cite{Tong:12,Zhou,ChatterjeeNoise}. One of the earliest works, focusing on FWM, demonstrated that FWM can increase the overall linewidth of the generated comb, both theoretically and experimentally \cite{Zhou}. In contrast, Tong et al. \cite{Tong:12} showed linewidth preservation over more than 200 nm in dual-pump, CW-seeded (continuous wave) parametric combs by employing phase-correlated pumps.

Early demonstrations of fiber-based parametric combs built entirely with polarization-maintaining (PM) components and PM highly nonlinear fiber (HNLF) achieved sub-40 kHz linewidths over ~100 nm bandwidth \cite{Ronit}. However, these studies also reported a “drastic increase of linewidth towards the edge of the comb bandwidth,” without identifying the specific noise sources responsible.

More recently, \cite{Yijia_apl} presented an all-PM parametric optical frequency comb (OFC) generator seeded by an EO comb, with a nonlinear amplifying loop mirror (NALM) acting as a pulse shaper. This system produced a 25-GHz-spaced comb with <10 kHz linewidth across 110 nm. Crucially, they showed that when driven by an ultra-low-noise RF source, the linewidth of the comb lines was dominated by the seed CW laser. However, when replaced with a higher-phase noise RF source, the linewidth increased. These results highlight the critical role of the RF source’s phase noise in determining the performance of parametric comb generation.

Despite the investigations discussed above \cite{Tong:12,ChatterjeeNoise,Zhou}, it remains unclear how the phase noise of a seeding electro-optic (EO) comb is transferred—or transformed—into the spectrally broadened comb after undergoing nonlinear processes. Specifically, it is well established that a standard EO comb follows the standard phase noise model \cite{Ishizawa,Heeboll:24}, in which the total phase noise, of a comb-line, is determined by two contributions: (1) common mode phase noise from the seed CW laser, and (2) repetition rate phase noise from the RF source. The open question is whether this standard phase model remains valid for nonlinearly broadened combs. In other words, do nonlinear processes introduce additional phase noise terms, or do they amplify the existing common mode and repetition rate noise?

Generally speaking, the answer to the aforementioned questions may depend on the type of nonlinear processes employed to broaden the initial comb. In this paper, we focus our investigation on the parametric comb reported in \cite{Yijia_apl}. Our investigation is based on a recently developed phase noise measurement technique that employs subspace tracking to identify various phase noise terms, and their scalings, as a function of comb-line number.

The paper is organized as follows: In section \ref{sec:gen phase noise model}, we briefly present the standard phase noise model, and discuss its extension for a nonlinearly broadened comb; In Section \ref{sec:experimental}, we describe the experimental setup and offline phase estimation technique; In Section \ref{sec:results}, we outline the principles of subspace tracking, introduce a framework for identifying the transformation of the phase noise model during nonlinear broadening, and present the corresponding results.

\section{Transformation of the standard phase noise model during nonlinear broadening}\label{sec:gen phase noise model}

According to the standard phase noise model, it is generally understood that the phase noise of an $m^{th}$ comb-line, $\phi_m(t)$, can be expressed as a combination of two contributions:

\begin{equation}\label{eq:std_phase_noise_model}
\phi_m(t) = \phi_{cm}(t) + m\phi_{rep}(t)
\end{equation}

where $\phi_{cm}(t)$ represents the common mode phase noise and $\phi_{rep}(t)$ is the repetition rate phase noise, and $m$ denotes the comb-line index.

However, as discussed in earlier theoretical works \cite{Benkler:05, Paschotta} and first experimentally demonstrated in \cite{Heeboll:24}, the standard phase noise model described by Eq.~(\ref{eq:std_phase_noise_model}) does not always hold. In such cases, additional phase noise contributions must be considered, leading to a generalized expression:

\begin{equation}\label{eq:elastic_tape}
\phi_m(t) = \phi_{cm}(t) + m\phi_{rep}(t) + \phi_{res}(m,t)
\end{equation}

Here, $\phi_{res}(m,t)$ represents the residual phase noise terms, which aggregate all additional noise sources beyond the standard model. Unlike the common mode and repetition rate terms, these residual contributions often lack a simple mathematical description or clear scaling behavior with the comb-line index.

For an electro-optic comb, it has been shown theoretically and experimentally that the standard phase noise model holds \cite{Ishizawa}:  

\begin{equation}\label{eq:EO comb phase noise model}
    \phi^{EO}_m(t) = \phi_{CW}(t) + m\phi_{RF}(t)
\end{equation}

where $m=-(M^{EO}-1)/2 ,...,(M^{EO}-1)/2$, and $M^{EO}$ is the total number of EO comb-lines. $\phi_{CW}(t)$ is a common mode phase noise term originating from the seed CW laser, and $\phi_{RF}(t)$ is the repetition rate phase noise term originating from the RF driving source. 

From a strictly mathematical point of view, the phase noise of the nonlinearly expanded frequency comb, $\phi_m^{Expan.}(t)$, is the result of a transformation of the phase noise of the input EO comb $\phi^{EO}_m(t)$. This can be expressed as: 

\begin{equation}\label{eq:PEO comb general}
    \phi^{Expan.}_m(t) = \mathcal{F}_{param.}\big(m,\phi_{CW}(t),\phi_{RF}(t)\big)
\end{equation}

where $\mathcal{F}_{param.}(\cdot)$ represents the transformation of the input common mode and repetition rate phase noise due to nonlinear processes. Typically,  $\mathcal{F}_{param.}(\cdot)$ will depend on the physical properties of the medium used to perform the spectral broadening, i.e.~dispersion profile, nonlinear coefficient, etc. The exact transformation performed by $\mathcal{F}_{param.}(\cdot)$ may therefore depend on implementation. Finally, to the best of the authors' knowledge, the exact expression for $\mathcal{F}_{param.}(\cdot)$ is typically not known, and we also anticipate that it is hard to derive. \vspace{0.1cm}     

However, using some physics intuition and relying on some already published results \cite{Zhou, Tong:12}, we can try to deduct how $\mathcal{F}_{param.}(\cdot)$ may transform the phase noise of the input EO comb. For instance, it has been shown in \cite{Zhou}, that FWM can lead to an increase in the total linewidth compared to the input linewidth, which implies the multiplication of the input phase noise terms. Additionally, nonlinear processes can also shift the comb-line that has the minimum phase noise \cite{lei2022optical}. For instance, for an EO comb, the comb-line with the minimum phase noise corresponds to $m=0$ according to Eq.~(\ref{eq:std_phase_noise_model}). Finally, nonlinear processes may also induce extra phase noise terms breaking the standard phase noise model. Taking the aforementioned into consideration, we proposed a transformed phase noise model for the nonlinearly broadened EO comb:     

\begin{equation}\label{eq: generalized phase noise model}
    \phi^{Expan.}_m(t) = N_0\phi_{CW}(t) + N_1(m-m^{\ast})\phi_{RF}(t) + \phi^{(1)}_{NL}(m,t)+...+\phi^{(p)}_{NL}(m,t)
\end{equation}

where $m=-(M^{Expan.}-1)/2,...,-(M^{EO}-1)/2,...,(M^{EO}-1)/2,...,(M^{Expan.}-1)/2$, and $M^{Expan.}$ is the total number of spectrally broadened comb-lines. The transformed phase noise model of the nonlinearly broadened comb expressed by Eq.~(\ref{eq: generalized phase noise model}), takes into account that nonlinear processes can multiply the common mode and the repetition rate phase noise by a constant factor $N_0$ and $N_1$, as well as shift the minimum phase noise comb-line corresponding to $m=m^{\ast}$. Finally, we assume that nonlinear processes can add extra $p$ phase noise terms that are functions of $\phi_{CW}(t)$ and $\phi_{RF}(t)$, i.e.~$\phi^{(i)}_{NL}(m,t)$, for $i=1,..,p$. The complete characterization of the phase noise performance of the comb expanded by nonlinear phenomena requires the identification and measurement of $N_0, N_1, m^{\ast}$ and $\phi^{(i)}_{NL}(m,t)$. 

In the following sections, we build an experimental setup and describe a method for identification of the transformed standard phase noise model of a nonlinearly expanded electro-optic frequency comb.   

\begin{figure}[t!]
    \centering
    \includegraphics[width=1\linewidth]{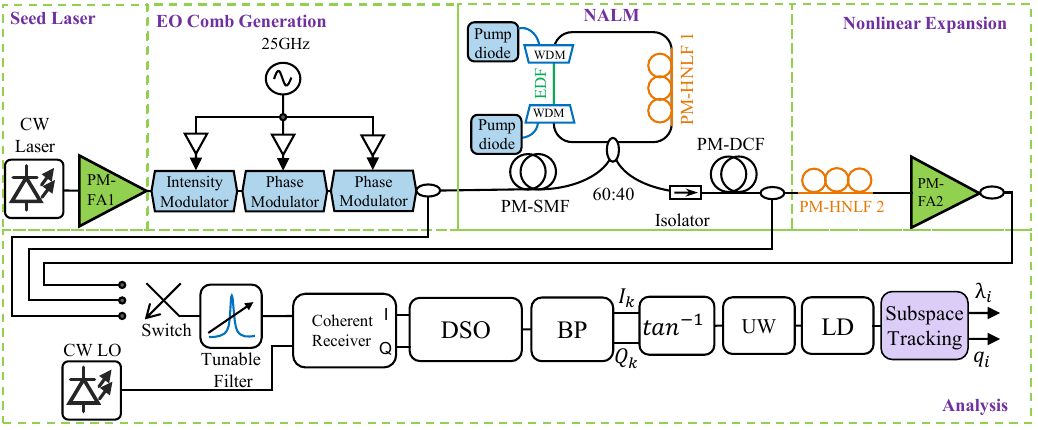}
    \caption{Experimental setup for generation and characterization of parametric optical frequency comb. DSO - Digital Signal Oscilloscope, BP - Bandpass Filter, UW - Unwrapping, LD - Linear Detrending.}
    \label{fig:setup}
\end{figure}

\section{Experimental setup}\label{sec:experimental}

The parametric optical frequency comb generator is shown in Fig.~\ref{fig:setup}. It consists of an electro-optic comb, a pulse-compression and -shaping module, and a nonlinear broadening stage. To measure the phase noise of the nonlinearly broadened frequency comb and test the transformed standard phase noise model, described in Section \ref{sec:gen phase noise model}, we employ a coherent detection, followed by a digital sampling scope to convert the downconverted comb-lines signal into the digital domain. \vspace{0.1cm} 

\begin{figure}[h!]
    \centering
    \includegraphics[width=0.75\linewidth]{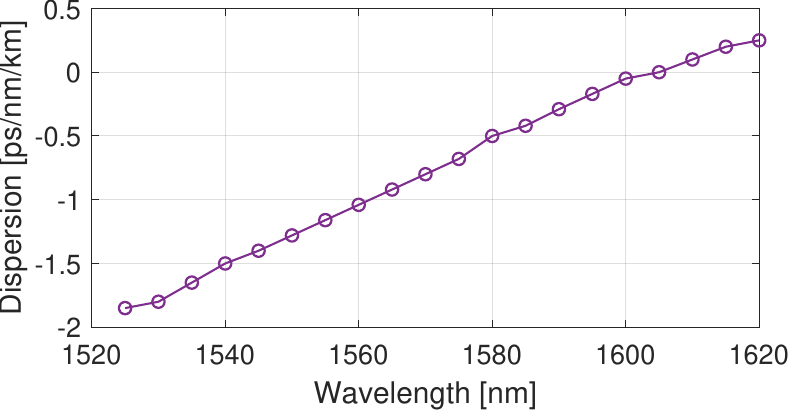}
    \caption{Dispersion profile of the PM-HNLF2.}
    \label{fig:dispersion}
\end{figure}

The frequency comb generator is entirely polarization-maintaining (PM) and operates without a resonant cavity, see Fig. \ref{fig:setup}. We employ a CW laser at 1550 nm, with a linewidth of $\sim$100 kHz, as a seed laser. The seed laser is first amplified by a PM fiber amplifier (PM-FA1). The amplified CW light is then sent through cascaded intensity and phase modulators, all driven by phase-synchronized 25 GHz RF signals from the same synthesizer, thereby creating an EO comb with 25 GHz spacing comprising approximately 76 tones. In the time domain, this corresponds to a chirped pulse train at 25 GHz repetition rate. These chirped pulses are compressed toward their transform limit by propagating them through 48 m of PM single-mode fiber (chromatic dispersion of $\approx$18 ps/nm·km at 1550 nm).

After the EO comb generation and pulse compression, short pulses of about 550 fs enter a nonlinear amplifying loop mirror (NALM), acting as a pulse shaper to suppress low-power pedestals resulting from imperfectly linear chirp. The NALM comprises a 60:40 coupler, a bidirectional erbium-doped fiber amplifier (EDFA) built around a highly doped PM erbium-doped fiber (peak absorption 80 dB/m), and a 40-m PM highly nonlinear fiber (PM-HNLF1) with dispersion $\approx$ -0.5 ps/nm·km at 1550 nm and zero-dispersion wavelength at 1565 nm (nonlinear coefficient 10.5 W$^{-1}$·km$^{-1}$). The detailed description of the NALM and its optimization procedure can be found in \cite{Yijia_apl}.

\begin{figure}[t!]
    \centering
    \includegraphics[width=1\linewidth]{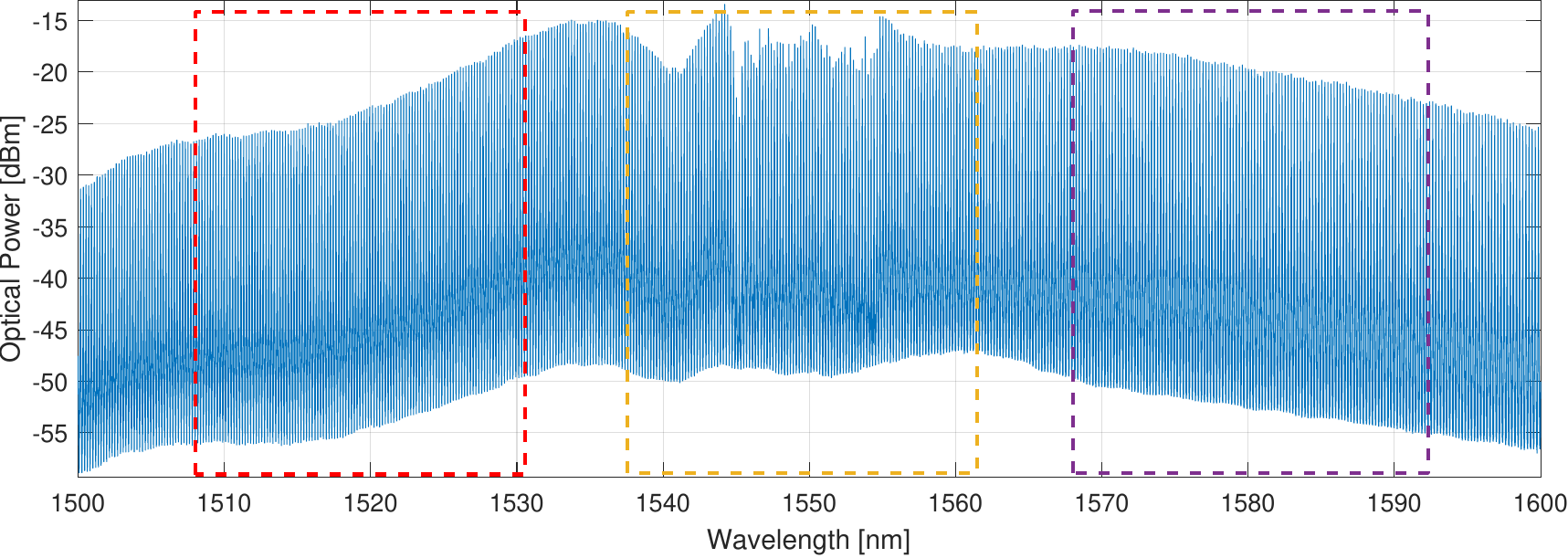}
    \caption{Optical frequency comb spectrum after the nonlinear broadening stage. Dashed boxes indicate the measured comb-lines.}
    \label{fig:spectr}
\end{figure}

Following the pulse shaping, the pulses are amplified again (PM-FA2) and launched into a second PM-HNLF for parametric spectral broadening. PM-HNLF2 has a length of 9 m and was chosen to have a normal dispersion of approximately -1.3 ps/nm·km and zero-dispersion wavelength near 1605 nm. The dispersion profile of PM-HNLF2 is shown in Fig.~\ref{fig:dispersion}. A PM dispersion-compensation fiber (DCF) is inserted before the PM-HNLF2 fiber to pre-compensate any residual dispersion from amplifiers and fiber pigtails, ensuring maximum peak power at the nonlinear mixer input. After nonlinear broadening, the output spectrum spans 1500–1610 nm with >0 dBm per tone. Spectral flatness and OSNR are monitored with an optical spectrum analyzer at 0.02 nm resolution. The optical spectrum after the nonlinear broadening stage is shown in Fig.~\ref{fig:spectr}.

To quantify how parametric nonlinear processes impact the phase noise of the spectrally broadened frequency comb, we analyze the signal after each successive stage, i.e.~EO comb generation, NALM pulse-shaping and nonlinear broadening. At each stage, we filter consecutive groups of five comb-lines from the regions indicated by the dashed lines in Fig.~\ref{fig:setup}. The filtering is performed using a tunable optical filter with a 0.95 nm bandwidth (EXFO XTA-50). After the EO comb and NALM output stages, the filtering is performed only around 1550 nm. After the nonlinear broadening stage, the filtering is performed around three wavelengths: 1520, 1550, and 1580 nm, see Fig.~\ref{fig:spectr}.

\begin{figure}[t!]
    \centering
    \includegraphics[width=1\linewidth]{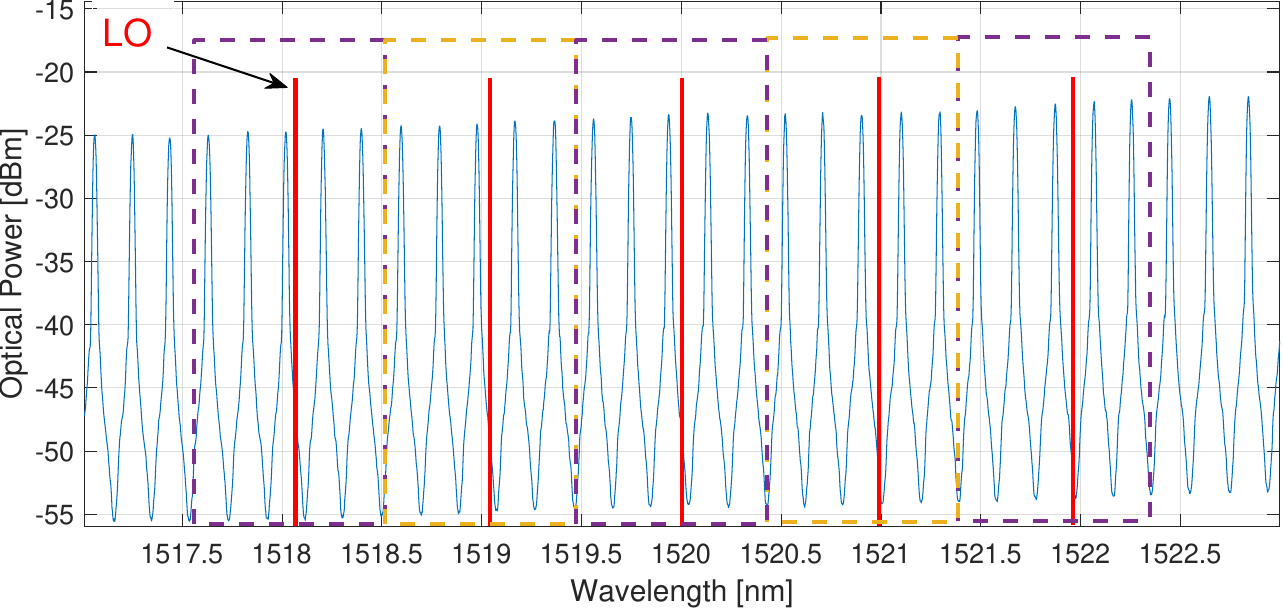}
    \caption{Illustration of the heterodyne measurement using groups of 5 comb-lines.}
    \label{fig:spectr_5lines}
\end{figure}

We move the filter center in discrete steps equivalent to 5 comb-line spacings, both toward longer and shorter wavelengths. The reason for detecting only 5 comb-lines, at a time, is due to the bandwidth limitations of the DSO, i.e.~100 GHz. As an example, from the initial 1520 nm setting, we took one measurement, then shifted the filter by ±5 comb-line spacings in each step, performing 12 steps toward higher wavelengths and 12 steps toward lower wavelengths, for a total of 25 measurements of 5 comb-lines around 1520 nm. This procedure samples the comb in blocks of 5 comb-lines, spaced so as to avoid overlap between successive groups, see Fig. \ref{fig:spectr_5lines}. We repeated the same measurement grid centered at 1550 nm and 1580 nm, each time stepping in increments of 5 comb-lines in both directions, to cover the spectral regions of interest uniformly. In total, this yields three measurement sweeps (around 1520 nm, 1550 nm, and 1580 nm), each comprising 25 measurements of 5 comb-lines, ensuring dense and systematic coverage of the comb spectrum for phase noise analysis. The measured spectral regions are depicted in Fig. \ref{fig:spectr} by dashed rectangular lines.

The filtered part of the spectrum is then coherently detected using a CW, widely tunable, local oscillator laser (CW LO), see Fig. \ref{fig:setup}. The employed CW LO laser is EXFO T100S. A polarization controller ensured that the optical power was aligned in the measured polarization state. The signal was then detected using dual quadrature coherent receiver (Kylia dual polarization 90 degree hybrid and $\sim$100 GHz balanced photodetectors), and finally sampled using the digital sampling oscilloscope (DSO) with the $F_s=256$ GSa/s sampling rate (Keysight UXR B-serials). The signal processing and the subsequent phase noise analysis are then performed offline.

\subsection{Digital signal processing based phase noise estimation}

For offline processing, we have access to the sampled in-phase and quadrature  signal components, i.e.~$I_k$ and $Q_k$, respectively. $k$ is a discrete time index, $k=1,...,K$ where $K$ is the total number of stored samples. Here, we used $K=32$ MSa. To extract the phase noise of the detected comb-lines, we first apply a bank of bandpass filters (BP), each with a 200 MHz bandwidth, to isolate individual frequency components from the detected heterodyne signal.

Since a dual quadrature coherent receiver is used as a detector, we can fully reconstruct the field in the digital domain and extract the phase noise by using the arctangent procedure $\tan^{-1}(I_k/Q_k)$. To eliminate discontinuities, we apply a phase unwrapping algorithm (UW) that corrects jumps greater than $\pi$ by adding or subtracting multiples of $2\pi$, resulting in a continuous phase evolution. After unwrapping, a linear detrending (LD) step is performed to isolate the phase fluctuations, providing the final phase noise traces, see Fig.~\ref{fig:setup}.



The extracted phase noise traces from each measurement group were then used to construct a sample covariance matrix:

\begin{equation}\label{eq:S_k}
\mathbf{S}(K) = \frac{1}{K-1} \sum_{k=1}^K \boldsymbol{\phi}_m(k)\bigg(\boldsymbol{\phi}_m(k)\bigg)^T \end{equation}

where ${\phi}_m(k)$ is the vector of measured phase noise of the comb-line $m$, and $t$ has been replaced by $k$, since we only have access to sampled signal, i.e.~$t=kT_s=k/Fs$. In total, we process all 25 groups of 5 comb-lines measured around each of the three spectral regions (1520, 1550, and 1580 nm), thereby covering more than three hundred comb-lines across the broadened spectrum. Eigenvalue decomposition of $\mathbf{S}(K)$ provides the eigenvectors and eigenvalues required for subspace analysis, which in turn allows us to identify the phase noise terms and their scaling behavior according to Eq.~(\ref{eq: generalized phase noise model}).



\section{Results and Discussion}\label{sec:results}

We assumed the standard "elastic-tape" phase noise model after the nonlinear broadening changes according to Eq. \ref{eq: generalized phase noise model}.  The identification of the transformed model then reduces to the identification of parameters $N_0$, $N_1$, and $m^*$, as well as potential extra phase noise terms $\phi^{(i)}_{NL}(m,t)$.

The parameters are most conveniently identified through a multi-heterodyne measurement performed around the comb center, in our case 1550~nm, because the wavelength of the seed laser corresponds to the comb-line $m=0$.
However, due to the large power variation within the original EO comb bandwidth after nonlinear expansion, phase noise measurement at 1550~nm was challenging~\cite{Yijia_apl, Ataie:14}. Therefore, we analyzed phase noise of the comb-lines around 1520 and 1580 nm.

\subsection{Subspace tracking}\label{sec:subspace}

For consistency, we provide a brief overview of the key ideas behind subspace tracking. A more detailed explanation can be found in~\cite{Razumov:23}. After signal detection, we extract the phase noise for each of the \(M\) detected comb-lines, denoted as \(\boldsymbol{\phi_m}(t)\). Each element of \(\boldsymbol{\phi_m}(t)\) represents the total phase noise of a comb-line, which can be modeled as a linear combination of common mode phase noise, repetition rate phase noise, and residual phase noise, as in Eq. (\ref{eq:elastic_tape}). In matrix form, this relationship is expressed as:

\begin{equation}
    \boldsymbol{\phi}_m(t) = \mathbf{H}\,\boldsymbol{\phi}^s(t),
\end{equation}

where $\boldsymbol{\phi}^s(t)$ contains phase noise terms, i.e. $[\phi_{cm}(t),\phi_{rep}(t),\phi_{res}(m,t)]$, and matrix \(\mathbf{H}\) is a generation matrix that captures how each phase noise term contributes to the total phase noise in the detected comb-lines. For instance, for the EO comb, which acts as the seed to the broadening stage, \(\mathbf{H}\) takes the following form:

\begin{equation}
    \mathbf{H} = 
    \begin{bmatrix}
        1 & -\tfrac{M-1}{2} \\
        \vdots & \vdots \\
        1 & \tfrac{M-1}{2}
    \end{bmatrix}
    \label{H}
\end{equation}

According to~\eqref{H}, the common mode phase noise term \(\phi_{cm}(t)\) contributes equally to each line, while the repetition rate term \(\phi_{rep}(t)\) contribution scales linearly across the comb-lines, directly following Eq. (\ref{eq:EO comb phase noise model}). 

The objective of subspace tracking is to estimate the underlying phase noise terms \(\boldsymbol{\phi}^s(t)\). Once \(\boldsymbol{\phi}^s(t)\) is obtained, one can compute the power spectral densities (PSDs) of the common mode, repetition rate, and residual phase noise terms. Obtaining \(\boldsymbol{\phi}^s(t)\) is achieved by determining a projection matrix \(\mathbf{G} \in \mathbb{R}^{P \times M}\) such that:

\begin{equation}\label{projection_eq}
    \boldsymbol{\phi}^s(t) = \mathbf{G}\,\boldsymbol{\phi}_m(t).
\end{equation}

As shown in~\cite{Razumov:23}, a valid choice for \(\mathbf{G}\) is given by:

\begin{equation}\label{eq:eigenvetors}
    \mathbf{G} = \mathbf{Q}_P^T,
\end{equation}

where \(\mathbf{Q}_P\) contains the \(P\) leading eigenvectors of the covariance matrix \(\mathbf{S} \in \mathbb{R}^{M \times M}\) of the measured phase noise \(\boldsymbol{\phi_m}(t)\), see Eq. (\ref{eq:S_k}). This projection isolates the subspace dominated by significant phase noise sources, enabling the estimation of \(\boldsymbol{\phi}^s(t)\).

The sample covariance matrix $\mathbf{S(K)}$ (Eq.~(\ref{eq:S_k})) can be decomposed as $\mathbf{S} = \mathbf{Q} \Lambda \mathbf{Q}^T$,
where $\Lambda$ is a diagonal matrix containing the eigenvalues, and $Q$ is a matrix whose columns are the corresponding eigenvectors. In the context of an optical frequency comb, these eigencomponents have direct physical meaning~\cite{Razumov:23}. The eigenvalues evolution corresponds to the variance evolution of distinct phase noise terms - namely, the common mode phase noise $\phi_{\text{cm}}(t)$, the repetition rate phase noise $\phi_{\text{rep}}(t)$, and residual noise $\phi_{\text{res}}(m,t)$. The eigenvectors indicate how these noise terms contribute to the total phase noise in the measured comb-lines.

By projecting the measured phase noise traces $\phi_m(t)$ onto the eigenvectors, one can isolate the individual contributions $\phi_{\text{cm}}(t)$,  $\phi_{\text{rep}}(t)$, and $\phi_{\text{res}}(m,t)$, as described in Eq.~(31) of~\cite{Razumov:23}. The corresponding power spectral densities (PSDs) of these components are computed using: $S_{\phi}^{cm/rep/res}(f) = \sum_{n}[\frac{1}{T}|FFT(\phi^{cm/rep/res}(t))T_s|^2]/N$,
where \( T \) is the total observation time, $T_s$ is the sampling interval, and $N$ is the number of phase noise traces used for averaging. For the given experiment, we used $N=5$ independent traces for the PSD averaging. 

\begin{figure}[t!]
    \centering
    \includegraphics[width=\linewidth]{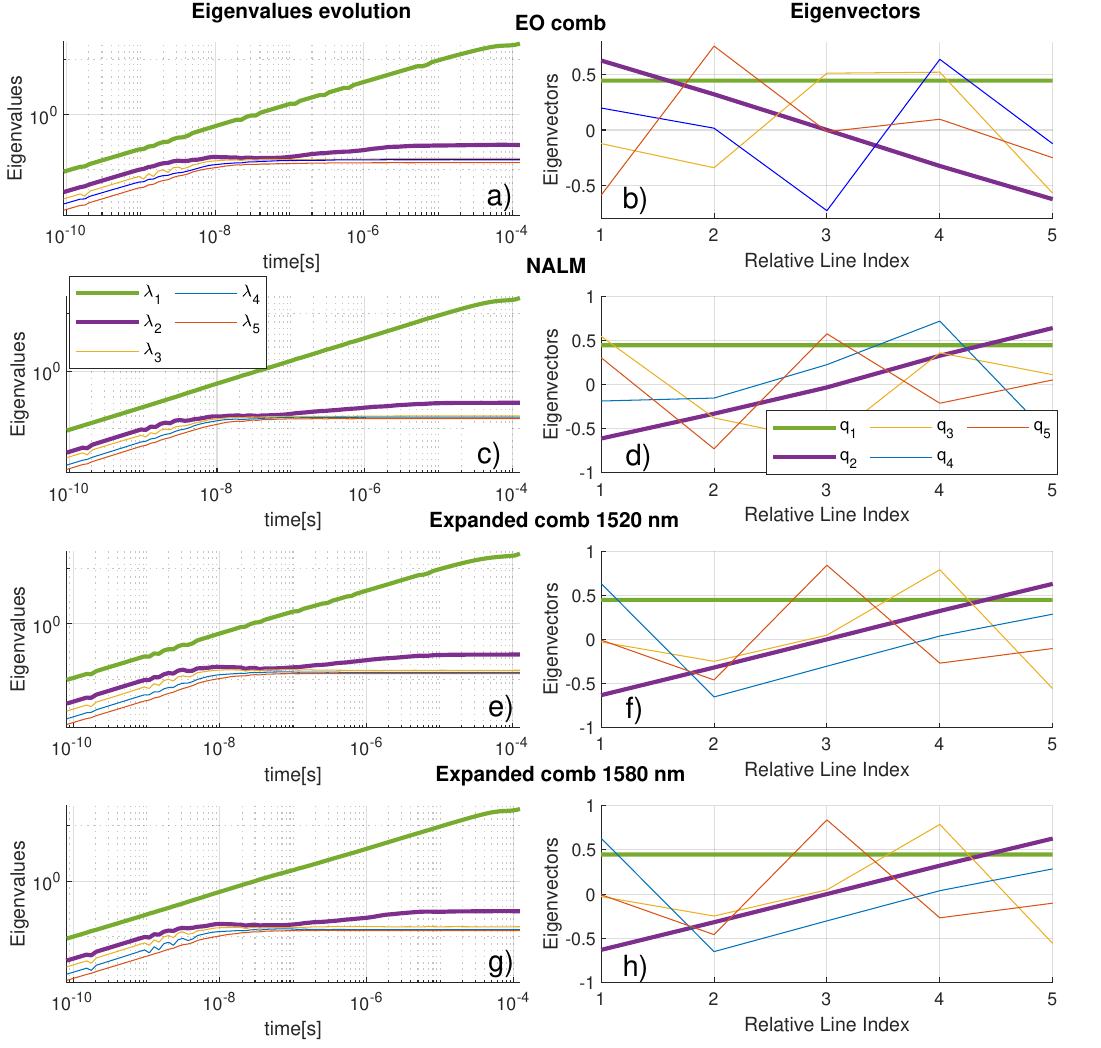}
    \caption{Evolution of eigenvalues and eigenvectors for 5 measured comb-lines after each parametric stage: (a–b) EO comb; (c–d) NALM; (e–f) expanded comb at 1520 nm; (g–h) expanded comb at 1580 nm.}
    \label{fig:eigva_eigve}
\end{figure}

\subsection{Identification of the extra phase noise terms $\phi^{(i)}_{NL}(m,t)$}

Fig.~\ref{fig:eigva_eigve} depicts the evolution of eigenvalues and eigenvectors (Eqs.~(\ref{projection_eq}) and (\ref{eq:eigenvetors})) for 5 measured comb-lines after each stage of nonlinear broadening. All 25 groups of 5 comb-lines measurements yield the same eigenvalues and eigenvectors, and at each stage, the eigenvalue and eigenvector plots resemble each other. From the eigenvalue evolution (Fig.~\ref{fig:eigva_eigve} a,c,e,g), we see $\lambda_1$ and $\lambda_2$ growing over time, while other eigenvalues $\lambda_3-\lambda_5$ remain constant, indicating the measurement noise floor\cite{Razumov:23}. The eigenvectors $\mathbf{q_1}$ and $\mathbf{q_2}$, associated with $\lambda_1$ and $\lambda_2$, are flat and linear, respectively, indicating that the first phase noise source (common mode, $\phi_{\text{cm}}(t)$) contributes equally to all comb-lines, while the second source (in this case RF source, $\phi_{\text{rep}}(t)$) contributes linearly across comb-line index. Other eigenvectors ($\mathbf{q_3}-\mathbf{q_5}$) are random and do not convey any useful information, consistently with the behavior of their eigenvalues. The observed behavior is standard for the EO combs \cite{Parriaux:20,Razumov:23}, where phase noise is usually described by two independent phase noise terms - common mode and RF source phase noise. The fact that the behavior of eigenvalues and eigenvectors is standard for EO combs, and that this behavior is preserved after every stage, demonstrates that nonlinear broadening does not introduce additional phase noise sources. This conclusion holds within our measurement sensitivity, the explored offset-frequency range ($\ge8$ kHz), and the specific experimental conditions (PM components, chosen HNLF, pump powers, etc.). The residual terms $\phi_{\text{res}}(m,t)$ are attributed solely to the measurement noise because $\lambda_3-\lambda_5$ are constant, and $\mathbf{q_3}-\mathbf{q_5}$ are random\cite{Razumov:23}. 

Having established that parametric generation does not introduce additional phase noise terms, i.e. $\phi^{(i)}(m,t)=0$, the next step is to identify the parameters $N_0$, $N_1$, and $m^*$ from Eq. (\ref{eq: generalized phase noise model}).

\subsection{Identification of the parameter $N_1$}

\begin{figure}[h!]
    \centering
    \includegraphics[width=0.75\linewidth]{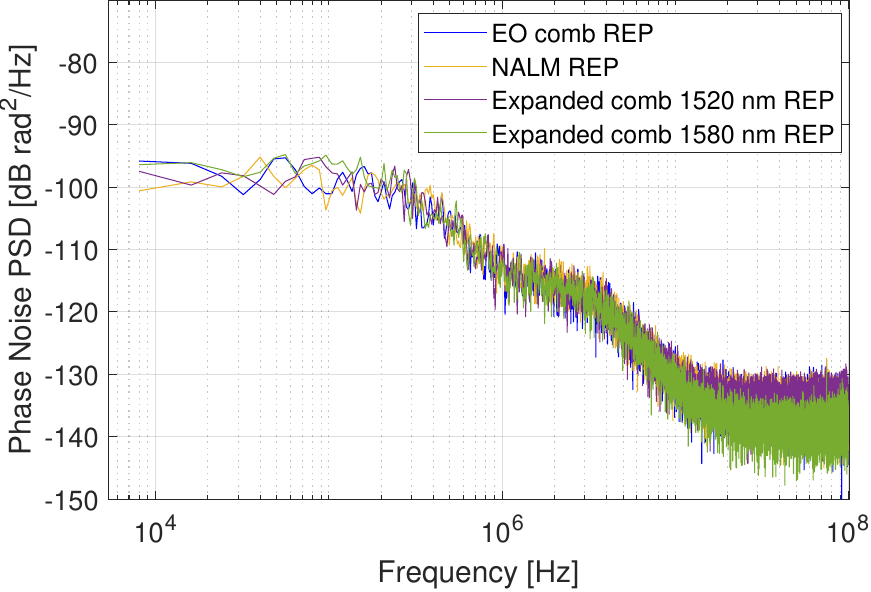}
    \caption{Power spectral densities of the repetition rate phase noise $\phi_{rep}(t)$ after each stage of nonlinear broadening.}
    \label{fig:psd_rr}
\end{figure}

First, we can identify $N_1$ by investigating the influence of the parametric generation process on the repetition rate term $\phi_{rep}(t)$. Fig.~\ref{fig:psd_rr} displays PSD of the repetition rate phase noise after each stage of nonlinear broadening, obtained using procedure described in Secs.~\ref{sec:experimental} and \ref{sec:subspace}. All 25 groups of 5 comb-lines measurements yield the same PSD curve. We observe an overlap between the PSDs after each of the stages. This indicates that parametric generation does not influence the repetition rate phase noise term, i.e. $N_1 = 1$. 

\subsection{Identification of the parameters $N_0$ and $m*$}

To determine $N_0$, we need first to note that subspace tracking yields phase noise components that are relative to the selected comb-lines. As a result, when applied to comb-lines located far from the comb center, the extracted common mode phase noise term $\phi_{\text{cm}}(t)$ appears overestimated. For instance, if subspace tracking is applied to comb-lines around 1520\,nm - where the central line-index under analysis is $m = 150$ - the estimated common mode phase noise becomes $\phi_{150}(t)$, i.e. phase noise of the comb-line $m=150$, not the actual $\phi_{\text{cm}}(t)$. Similarly,  comb-lines around 1580 nm correspond to the central line-index $m=-150$.

In order to identify $N_0$ and $m^*$, we need to analyze $\phi^{Expan.}_{\text{cm}}(t)$ of the expanded comb and compare it with $\phi^{EO}_{\text{cm}}(t)$ of the EO comb before broadening. Although measurements are only available at 1520 and 1580~nm, Fig.~\ref{fig:psd_rr} confirms that $\phi_{rep}(t)$ remains unchanged after broadening and is the same across all comb lines. Therefore, two equivalent approaches can be used to find $\phi^{Expan.}_{\text{cm}}(t)$.


In the first approach, instead of projecting the measured phase noise of individual comb-lines $ \boldsymbol{\phi}_m(t) $ onto the eigenvector basis \( \mathbf{q_1},\dots,\mathbf{q_5} \), we used the projection matrix $\mathbf{G}$ obtained via the Moore-Penrose pseudoinverse $\mathbf{G}=(\mathbf{H}^T\mathbf{H})^{-1}\mathbf{H}^T$, where the generation matrix \( \mathbf{H} \) is defined for the comb-lines with the central line-index $m=150$ as:

\begin{equation}
\mathbf{H} =
\begin{bmatrix}
1 & 148 \\
\vdots & \vdots \\
1 & 152
\end{bmatrix},
\label{H_corr}
\end{equation}

which corresponds to the indices of the five comb-lines under analysis (from \( m-2 \) to \( m+2 \)). This matrix relates the measured phase noise to the underlying contributions from \( \phi_{\text{cm}}(t) \) and \( \phi_{\text{rep}}(t) \). A similar construction applies when analyzing the region around 1580\,nm, but with \( m = -150 \).

\begin{figure}[h!]
    \centering
    \includegraphics[width=0.75\linewidth]{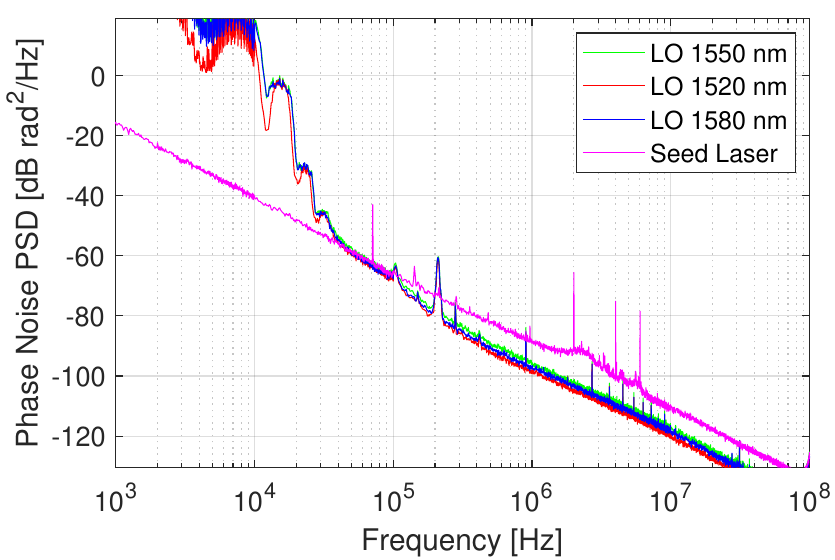}
    \caption{Phase noise power spectral densities of seed laser and LO at 1520, 1550, and 1580 nm obtained using OEWaves phase noise analyzer.}
    \label{fig:lo_seed}
\end{figure}

In the second approach, we first project \( \boldsymbol{\phi}_m(t) \) onto the eigenvectors \( \mathbf{q_1},\dots,\mathbf{q_5} \) to obtain the phase trace \( \phi_{150}(t) \), and then perform the scaling procedure from~\cite{Razumov:23}. Assuming that:

\begin{equation}\label{m*_assumption}
\phi_{150}(t) = \phi_{\text{cm}}(t) + 150 \times \phi_{\text{rep}}(t),
\end{equation}

we can isolate \( \phi_{\text{cm}}(t) \). Both methods yield consistent results.


Next, we estimate the power spectral density of $\phi^{Expan.}_{\text{cm}}(t)$ and compare it with $\phi^{EO}_{\text{cm}}(t)$ and $\phi^{NALM}_{\text{cm}}(t)$. Note that in a heterodyne measurement, the measured phase noise equals the sum of the device-under-test phase noise and the LO phase noise. Because the EO comb is used as a seed source, the common mode phase noise term is given by a sum $\phi_{\text{cm}}(t) = \phi_{\text{LO}}(t) + \phi_{\text{seed}}(t)$. Fig. \ref{fig:lo_seed} shows phase noise PSDs of the seed laser and LO at 1520, 1550, and 1580 nm measured using the OEwaves phase noise analyzer. We observe that below 40 kHz, the LO phase noise exceeds the seed laser phase noise, hence LO noise will dominate heterodyne measurements. Therefore, further we present PSD plots starting from 40 kHz, where LO and seed noise are comparable, or seed phase noise dominates. 

\begin{figure}[h]
    \centering
    \includegraphics[width=\linewidth]{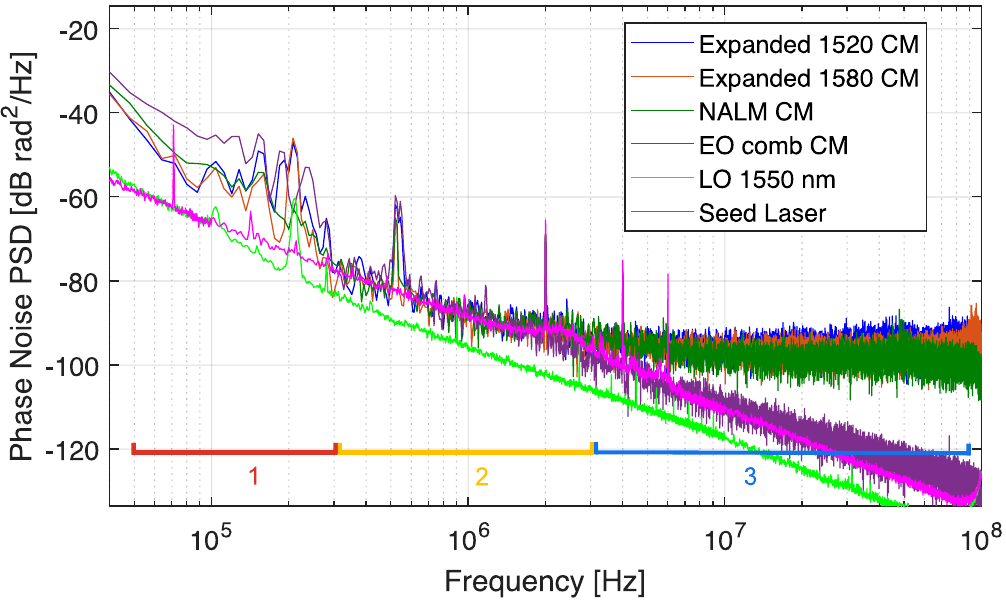}
    \caption{Power spectral densities of the common mode phase noise $\phi_{\text{cm}}(t)$ after each stage ofnonlinear broadening compared with seed and LO laser noise.}
    \label{fig:psd_cm}
\end{figure}

Fig.~\ref{fig:psd_cm} shows the PSDs of the common mode phase noise $\phi_{\text{cm}}(t)$ after each stage of parametric generation, along with the PSDs of the seed and LO laser phase noise. All 25 groups of 5 comb-lines measurements yield the same PSD curve. The phase noise measurement at 1550~nm was not feasible and is therefore not included in the figure.

To better understand the results, the figure can be split into three frequency regions:

\begin{itemize}
  \item Region 1 (40 kHz-300~kHz): seed and LO phase noise is of similar magnitude. In this region, common mode phase noise after all the parametric generation stages shows a similar behavior, but lies $\sim$10–15 dB $rad^2$/Hz above both LO and seed phase noise PSDs.
  \item Region 2 (300~kHz–3~MHz): there the seed laser phase noise is higher than the LO phase noise. The common mode phase noise after all the stages is nearly identical to the seed laser's curve. From this we can conclude that between 300 kHz and 3 MHz nonlinear broadening does not contribute with any additional phase noise. 
  \item Region 3 (above 3~MHz): here common mode PSD after nonlinear expansion and NALM is dominated by the measurement noise floor ($\sim$ -100 dB rad$^2$/Hz) and therefore does not provide much useful information. The different noise floor of the EO comb measurement is due to the different signal-to-noise ratio of the comb-lines before and after the broadening. 
\end{itemize}

In terms of identifying $N_0$ and $m^*$, Fig. \ref{fig:psd_cm} shows that $\phi^{Expan.}_{\text{cm}}(t)=\phi^{EO}_{\text{cm}}(t) = \phi^{NALM}_{\text{cm}}(t)$, therefore $N_0 = 1$. Furthermore, because the assumptions in Eq.~\ref{H_corr} and Eq.~\ref{m*_assumption} proved to be correct, we obtain $m^*=0$, indicating that the central comb-line is the same as before the broadening.

Thus, the identified parameters are $N_0=1,N_1=1,m^*=0$, and the extra phase noise terms $\phi^{(i)}_{NL}(m,t)=0$, which leads to the following phase noise model of the expanded comb:

\begin{equation*}
\begin{aligned}
        \phi_m(t) = 1\times\phi_{cm}(t) + 1\times(m-0)\phi_{rep}(t) + 0  = \\
        =\phi_{cm}(t) + m\phi_{rep}(t)
\end{aligned}
\end{equation*}


\begin{figure}[t]
    \centering
    \includegraphics[width=0.75\linewidth]{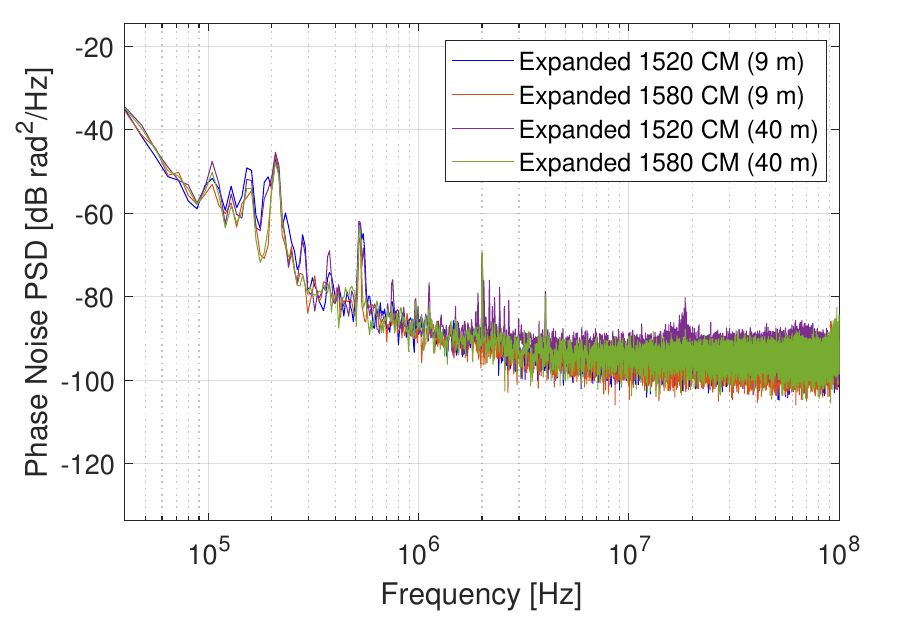}
    \caption{Comparison of the PSD of the common mode phase noise $\phi_{\text{cm}}(t)$ for 9 m vs. 40 m of HNLF at 1520 and 1580 nm.}
    \label{fig:long_fiber_cm}
\end{figure}

\subsection{Influence of HNLF length}

\begin{figure}[h]
    \centering
    \includegraphics[width=0.75\linewidth]{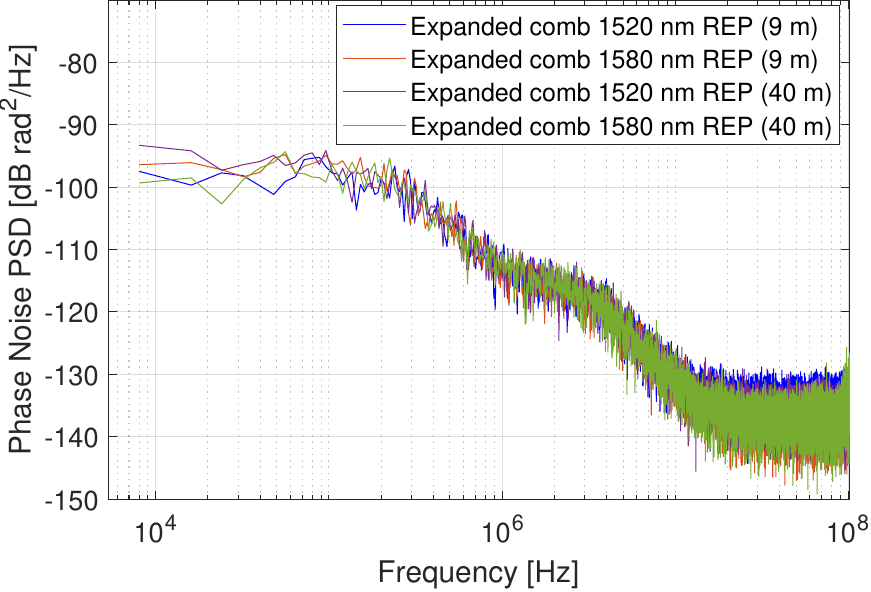}
    \caption{Comparison of the PSD of the repetition rate phase noise $\phi^{\text{rep}}(t)$ for 9 m vs. 40 m of HNLF at 1520 and 1580 nm.}
    \label{fig:long_fiber_rr}
\end{figure}

To further confirm our hypothesis, we conducted a phase noise investigation in which 9 meters of HNLF at the nonlinear expansion stage were replaced with 40 meters of a similar HNLF. The phase noise was then analyzed in terms of the common mode term $\phi_{\text{cm}}(t)$ and the repetition rate term $\phi^{\text{rep}}(t)$.

Figs. \ref{fig:long_fiber_cm} and \ref{fig:long_fiber_rr} compare PSDs of common mode phase noise $\phi_{\text{cm}}(t)$ and repetition rate phase noise $\phi^{\text{rep}}(t)$ for 9 m vs. 40 m of HNLF at 1520 and 1580 nm. The curves overlap, indicating that extending the HNLF length to 40 m does not significantly affect both the common mode phase noise and the repetition rate phase noise.

\begin{figure}[t]
    \centering
    \includegraphics[width=0.75\linewidth]{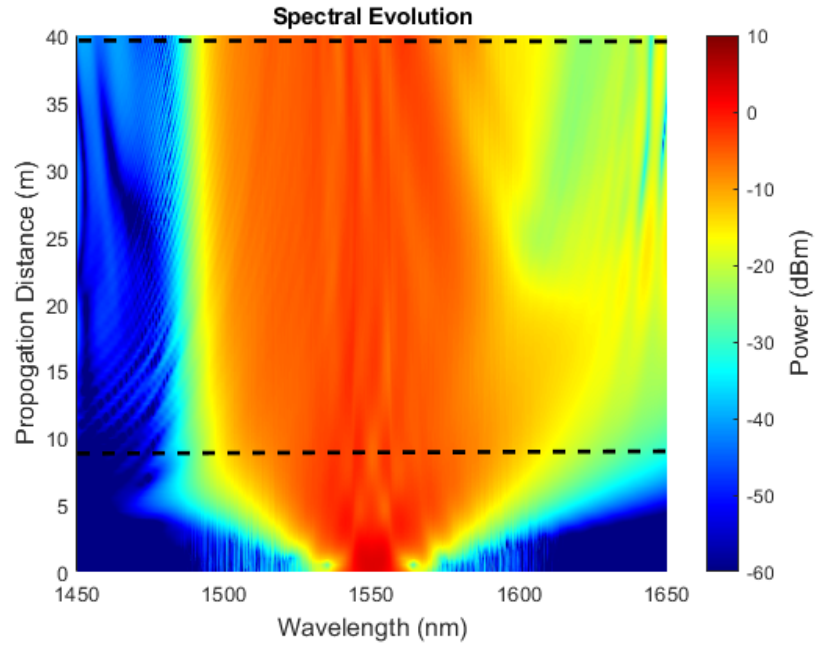}
    \caption{Simulated evolution of the frequency-comb spectrum as HNLF length increases from 0 m to 40 m.}
    \label{fig:spectral_evol}
\end{figure}

We support this investigation with simulations of the setup presented in Fig.~\ref{fig:setup}, sweeping HNLF length from 0 to 40 m at the nonlinear expansion stage. As shown in Fig.~\ref{fig:spectral_evol}, the simulated comb spectrum evolves up to ~9 m, beyond which bandwidth does not increase. Additional nonlinearity and dispersion beyond 9 m (up to 40 m) lead to decorrelation and fiber noise. However, Figs.~\ref{fig:long_fiber_cm} and~\ref{fig:long_fiber_rr} show that phase noise introduced by the extra 31 m is negligible in measured phase noise PSDs.

\section{Conclusion}
We have experimentally demonstrated that for the spectrally expanded comb under investigation, nonlinear processes do not generate any extra phase noise terms, nor do they amplify phase noise terms associated with the input comb. Additionally, we have shown that increasing the length of the nonlinear fiber at the broadening stage from 9 to 40 meters does not have an impact on the phase noise, even though spectral bandwidth saturates. The question which remains to be answered is whether the results of the investigation are general or do they apply only for the specific nonlinear expansion stage employed in this paper.  

\begin{backmatter}
\bmsection{Funding} Villum Fonden VI-POPCOM VIL54486 \& OPTIC-AI
VIL29334

\bmsection{Disclosures} The authors declare no conflicts of interest.

\bmsection{Data availability} Data underlying the results presented in this paper are not publicly available at this time but may be obtained from the authors upon reasonable request.

\end{backmatter}



\bibliography{sample}






\end{document}